\documentclass{PoS}
\usepackage{epstopdf}
\title{Infrared divergence of the color-Coulomb self-energy in Coulomb gauge QCD}

\ShortTitle{Infrared divergence of the color-Coulomb self-energy in Coulomb gauge QCD}

\author{\speaker{Y. Nakagawa}\\
Research Center for Nuclear Physics, Osaka University,\\
Ibarakisi, Osaka 567-0044, Japan\\
E-mail: \email{nkgw@rcnp.osaka-u.ac.jp}}

\author{T. Saito, H. Toki\\
Research Center for Nuclear Physics, Osaka University,\\
Ibarakisi, Osaka 567-0044, Japan\\
E-mail: \email{tsaito@rcnp.osaka-u.ac.jp},
\email{toki@rcnp.osaka-u.ac.jp}}

\author{A. Nakamura\\
Research Institute for Information Science and Education, Hiroshima University,\\
Higashi-Hiroshima 739-8521, Japan\\
E-mail: \email{nakamura@riise.hiroshima-u.ac.jp}}

\abstract{
We investigate the spectrum of the Faddeev-Popov operator in Coulomb gauge QCD using quenched SU(3) lattice simulation.
In the confinement phase, we observe the accumulation of the near-zero modes of the FP operator at large lattice volumes, and the color-Coulomb self-energy diverges in the infrared limit.
Moreover, even in the deconfinement phase, the behavior of the FP eigenvalue density is qualitatively the same as in the confinement phase and the color-Coulomb self-energy is infrared divergent.
}

\FullConference{XXIVth International Symposium on Lattice Field Theory\\

                July 23-28, 2006\\

                Tucson, Arizona, USA}

\begin{document}

\section{Introduction}

The confinement mechanism in Coulomb gauge QCD has received a lot of attention recently.
It is firstly discussed by Gribov in '70s that the instantaneous interaction provides the long range interaction \cite{GribovVN:NPB139:1978}, and this is further elaborated by Zwanziger recently \cite{ZwanzigerD:NPB518:1998}.
Zwanziger showed that the color-Coulomb potential which is the instantaneous interaction energy between heavy quarks is stronger than a physical potential.
This inequality tells us that the necessary condition for the physical potential being a confining potential is that the color-Coulomb potential is also a confining potential, i.e., "no confinement without color-Coulomb confinement" \cite{ZwanzigerD:PRL90:2003}.

The recent Monte Carlo simulations in the SU(2) and SU(3) lattice gauge theories showed that the color-Coulomb potential rises linearly with distance, and its string tension has $2\sim 3$ times larger value than that of the Wilson potential, which is an expected result from the Zwanziger's inequality.  \cite{GreensiteJ:PRD67:2003,NakamuraA:PTP115:2006}.
In addition, it was shown that an asymptotic scaling violation for the color-Coulomb string tension is weaker than that of the Wilson string tension \cite{GreensiteJ:PRD69:2004, NakagawaY:PRD73:2006}. Furthermore, it has been shown that the color-Coulomb potential is a confining potential even in the deconfinement phase \cite{GreensiteJ:PRD67:2003,NakamuraA:PTP115:2006}.
Thus the color-Coulomb string tension does not serve as an order parameter for the confinement/deconfinement phase transition.

In Gribov-Zwanziger confinement scenario, the near-zero modes of the Faddeev-Popov (FP) operator play a crucial role to produce the singular behavior of the ghost propagator in the infrared region.
The color-Coulomb potential in the color-singlet channel is given by
\begin{equation}\label{CCP}
V_c(\vec{x}-\vec{y})\equiv g^2\textrm{Tr}[T^aT^b]\left\langle\int d^3z\mathcal{G}^{ac}(\vec{x}, \vec{z}; A^{{tr}})(-\nabla_{\vec{z}}^2)\mathcal{G}^{cb}(\vec{z}, \vec{y}; A^{{tr}})\right\rangle,
\end{equation}
where $\mathcal{G}$ is the Green's function of the FP operator and $\langle\cdot\rangle$ denotes an Euclidean expectation value.
$T^a$ ($a=1,..., 8$) are the generators of $\mathfrak{su}(3)$ Lie algebra.
The singular behavior of the ghost propagator in the infrared region leads to the long-range interaction of the color-Coulomb potential which is responsible for the color confinement.

Recently Greensite, Olejn\'{i}k and Zwanziger discussed the self-energy of an isolated quark and derived the necessary condition for the color confinement \cite{GreensiteJ:JHEP05:2005}.
The authors studied the spectrum of the FP operator in Coulomb gauge using SU(2) lattice gauge simulation and confirmed that the necessary condition is satisfied.
In this study, we investigate the distribution of the FP eigenvalues in SU(3) lattice gauge simulations, and check whether the necessary condition for color confinement is satisfied or not.

\section{Color-Coulomb self-energy}

The Coulomb gauge Hamiltonian can be expressed as the sum of the transverse part and the instantaneous part \cite{CucchieriA:PRD65:2002}:
\begin{eqnarray}\label{Hamiltonian}
H
&=&\frac{1}{2}\int d^3x\left((E_i^{a,{tr}}(\vec{x},t))^2+B_i^a(\vec{x},t)^2\right)+\frac{1}{2}\int d^3y\int d^3z\rho^a(\vec{y}, t)\mathcal{V}^{ab}(\vec{y},\vec{z}; A^{{tr}})\rho^b(\vec{z}, t).
\end{eqnarray}
Here $E_i^{a,{tr}}$ are the transverse components of the color electric field, $B_i^a$ the color magnetic field, $\rho^a(\vec{x},t)$ the color charge density.
The kernel of the instantaneous interaction is given by
\begin{equation}\label{kernel}
\mathcal{V}^{ab}(\vec{y},\vec{z}; A^{{tr}})\equiv\int d^3x\mathcal{G}^{ac}(\vec{y}, \vec{x}; A^{{tr}})(-\nabla_{\vec{x}}^2)\mathcal{G}^{cb}(\vec{x}, \vec{z}; A^{{tr}}),
\end{equation}
where $A_i^{a,{tr}}$ are the transverse components of the gluon field.
$\mathcal{G}$ is the Green's function of the FP operator $M^{ab}=-\partial_iD_i^{ab}=-\delta^{ab}\partial^2-gf^{abc}A_i^{c,{tr}}\partial_i$.
On a lattice, the FP operator is an $8V_3\times 8V_3$ sparse matrix ($V_3$ is the lattice 3-volume) and expressed in terms of SU(3) spatial link variables $U_i$ as 
\begin{eqnarray}\label{FPop}
M^{ab}_{xy}=&&\sum_{i}\mathfrak{Re}\textrm{Tr}\left[\{T^a, T^b\}\left(U_{i}(x)+U_{i}(x-\hat{i})\right)\delta_{x, y}\right.\nonumber\\
&&\qquad\qquad \left.\frac{}{}-2T^bT^aU_{i}(x)\delta_{y, x+\hat{i}}-2T^aT^bU_{i}(x-\hat{i})\delta_{y, x-\hat{i}}\right].
\end{eqnarray}

The color-Coulomb self-energy for an isolated color charge, whose energy diverges in the infrared limit in a confining theory, is \cite{GreensiteJ:JHEP05:2005}
\begin{equation}\label{self-energy}
E_c=\textrm{Tr}[T^aT^b] g^2\langle \mathcal{V}^{ab}(\vec{x},\vec{x}; A^{\textrm{tr}})\rangle.
\end{equation}
The color-Coulomb self-energy is ultraviolet divergent in the continuum limit both in an abelian and a non-abelian gauge theories, and can be regularized by introducing the cutoff.
The interesting point is that the infrared divergence may exist in a confining theory at infinite volume.

We define the normalized density of the FP eigenvalues
\begin{equation}
\rho(\lambda)\equiv\frac{N(\lambda, \lambda+\Delta\lambda)}{8V_3\Delta\lambda},
\end{equation}
where $N(\lambda, \lambda+\Delta\lambda)$ is the number of eigenvalues in the range $[\lambda, \lambda+\Delta\lambda]$.
By expanding the Green's function of the FP operator in terms of the eigenvectors $\phi^a_n(\vec{x})$ and the eigenvalues $\lambda_n$ of the FP operator, we find
\begin{equation}
E_c=g^2C_D\int^{\lambda{max}}_0 d\lambda\frac{\left\langle\rho(\lambda)F(\lambda)\right\rangle}{\lambda^2},
\end{equation}
where $C_D (>0)$ is the Casimir invariant for the representation $D$ and the upper limit of the integration $\lambda_{\max}$ corresponds to the UV lattice cutoff.
In the Gribov-Zwanziger scenario the gauge configurations are restricted to the Gribov region, and therefore the lower limit of the integration is zero.
$F_n$ are the expectation values of the negative Laplacian in the FP eigenmodes,
\begin{equation}
F_n=\int d^3x\phi^{\ast a}_n(\vec{x})(-\nabla^2)\phi^a_n(\vec{x}).
\end{equation}
If the condition
\begin{equation}\label{condition}
\lim_{\lambda\to 0}\frac{\langle\rho(\lambda)F(\lambda)\rangle}{\lambda}>0
\end{equation}
is satisfied in the infinite volume limit, the color-Coulomb self-energy diverges in the infrared region.
This is the necessary condition for the color confinement \cite{GreensiteJ:JHEP05:2005}.

The FP eigenvalue density of the near-zero modes is closely related to the infrared behavior of the color-Coulomb potential.
From Eqs. (\ref{CCP}) and (\ref{self-energy}), the color-Coulomb self-energy can be expressed as
\begin{equation}\label{mspace}
E_c=V_c(\vec{x}-\vec{x})=\int\frac{d^3p}{(2\pi)^3}\tilde{V}_c(\vec{p})=\int_0^{\Lambda}\frac{d|\vec{p}|}{4\pi}|\vec{p}|^2\tilde{V}_c(|\vec{p}|).
\end{equation}
Here we introduce the ultraviolet cutoff $\Lambda$.
If the condition (\ref{condition}) is satisfied, the color-Coulomb self-energy diverges in the infrared region.
Accordingly, the right-hand side of (\ref{mspace}) diverges in the infrared limit.
It means that the color-Coulomb potential is more singular in the infrared region than the Coulomb potential $\tilde{V}(\vec{p})\sim 1/|\vec{p}|^2$.

The FP operator is the negative Laplacian in the abelian gauge theory.
Thus the FP eigenfunctions are the plane waves and $\lambda=\vec{k}^2$.
By counting the number of states in momentum space, it is easy to show that
\begin{equation}\label{abelian}
\rho(\lambda) = \frac{\sqrt{\lambda}}{4\pi^2},\qquad F(\lambda)=\lambda,
\end{equation}
in the infinite volume limit.
Obviously the necessary condition (\ref{condition}) is not satisfied in this case.

\section{Numerical simulations}

We calculate the FP eigenvalue density by the $SU(3)$ lattice gauge simulations in quenched approximation.
The lattice configurations are generated by the heat-bath Monte Carlo technique with the Wilson plaquette action.
In these simulations we adopt the iterative method to fix a gauge.
We used the ARPACK package to evaluate the lowest 1000 eigenvalues and corresponding eigenvectors of the FP operator.

\subsection{$\langle\rho(\lambda)F(\lambda)\rangle/\lambda$ in the confined phase}

\begin{figure}[htbp]
\begin{center}
\resizebox{7cm}{!}{\includegraphics{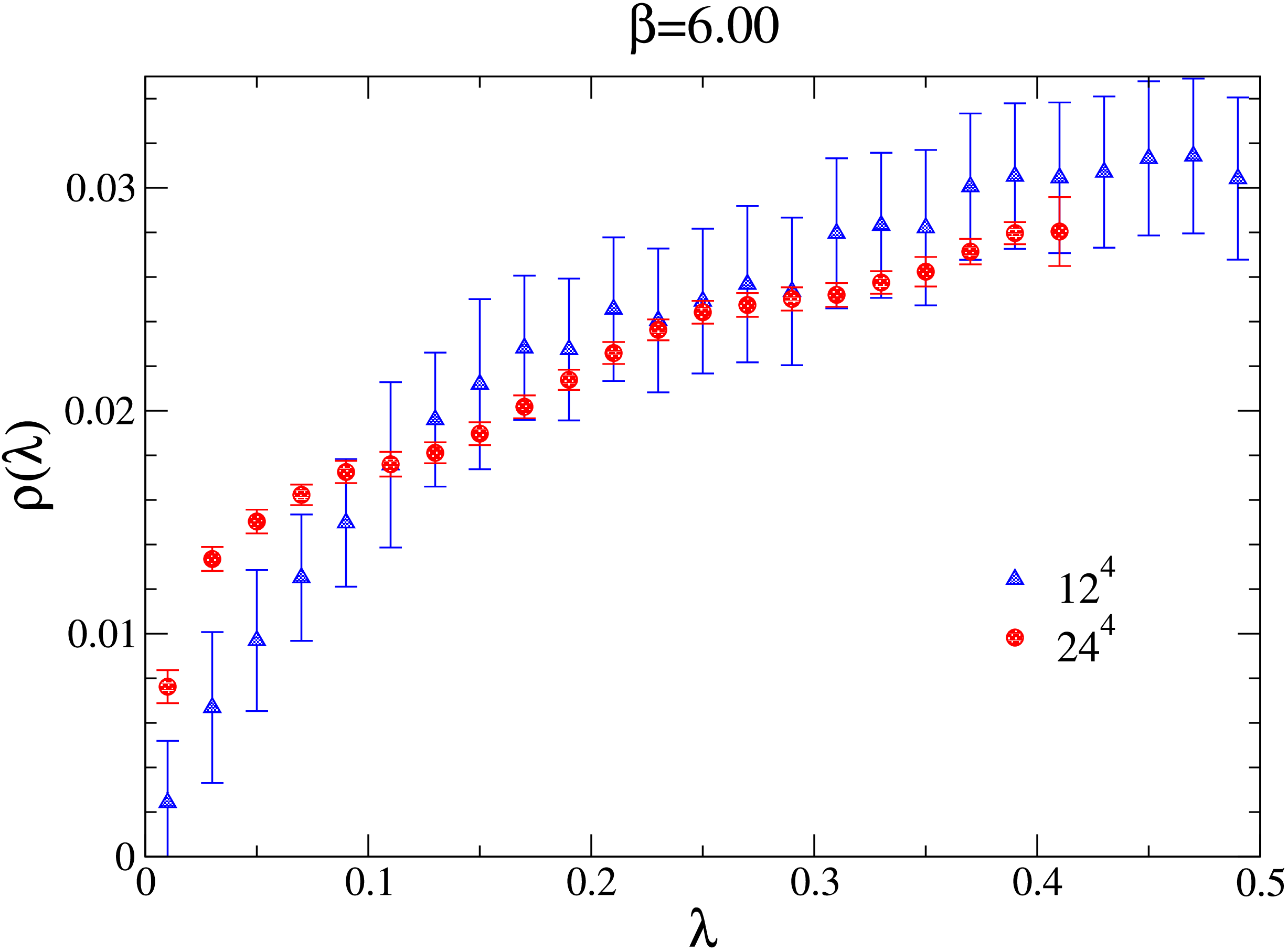}}
\resizebox{7cm}{!}{\includegraphics{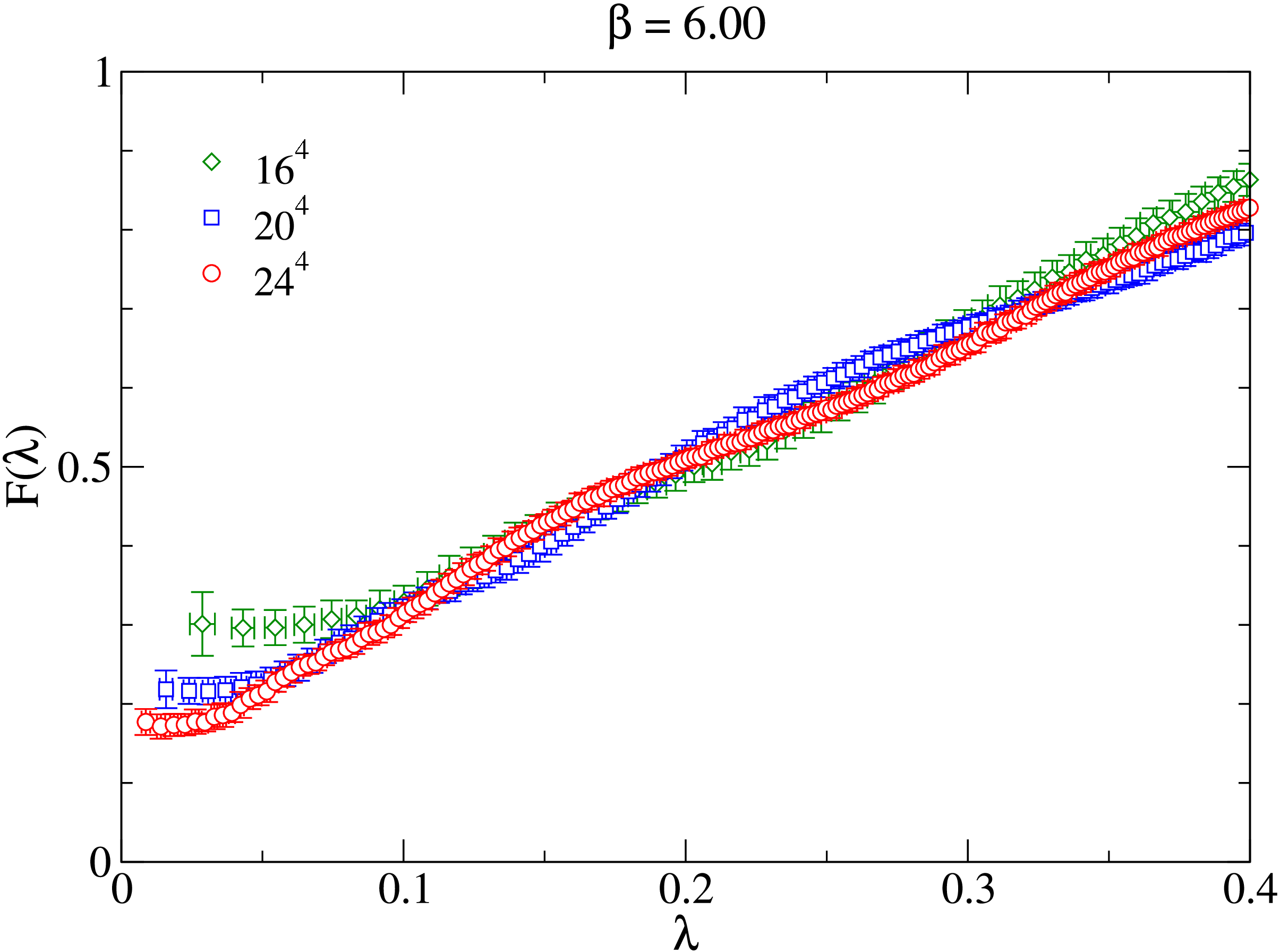}}
\caption{(a) The FP eigenvalue density in the confinement phase. (b) The average Laplacian $F(\lambda)$ in the confinement phase.}
\label{EVzero}
\end{center}
\end{figure}

Figures \ref{EVzero}(a) and (b) show $\rho(\lambda)$ and $F(\lambda)$ at $\beta=6.0$ on a variety of lattice sizes.
We see the accumulation of the near zero modes of the FP operator at larger lattice volume.
On the other hand, $\rho(\lambda)$ is almost saturated above $\lambda\sim 0.15$.
$F(\lambda)$ becomes flat at smaller value of $\lambda$ as the lattice volume increases and it seems that as $\lambda\to 0$ the average Laplacian approaches positive constant in the infinite volume limit.

\begin{figure}[htbp]
\begin{center}
\resizebox{7cm}{!}{\includegraphics{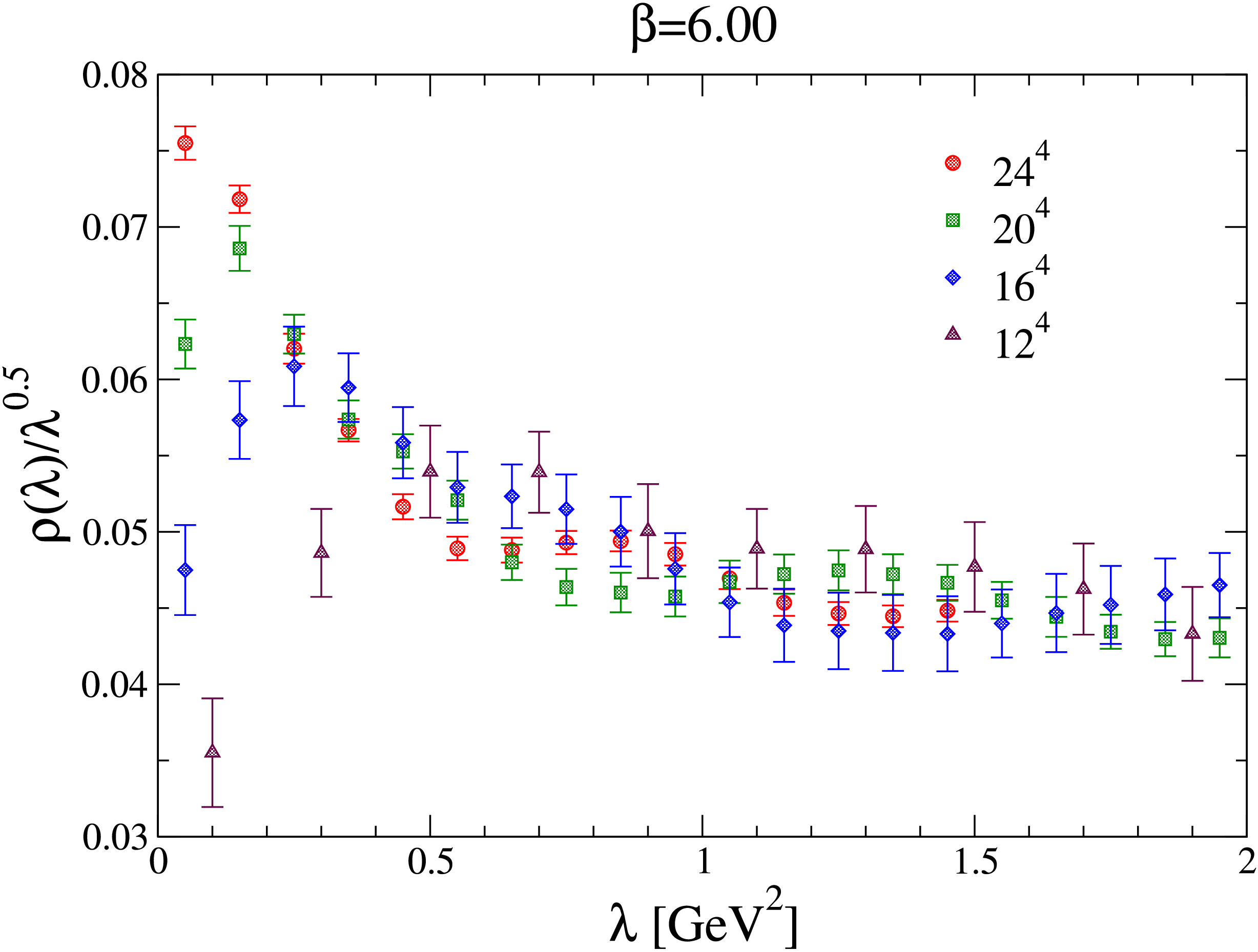}}
\resizebox{7cm}{!}{\includegraphics{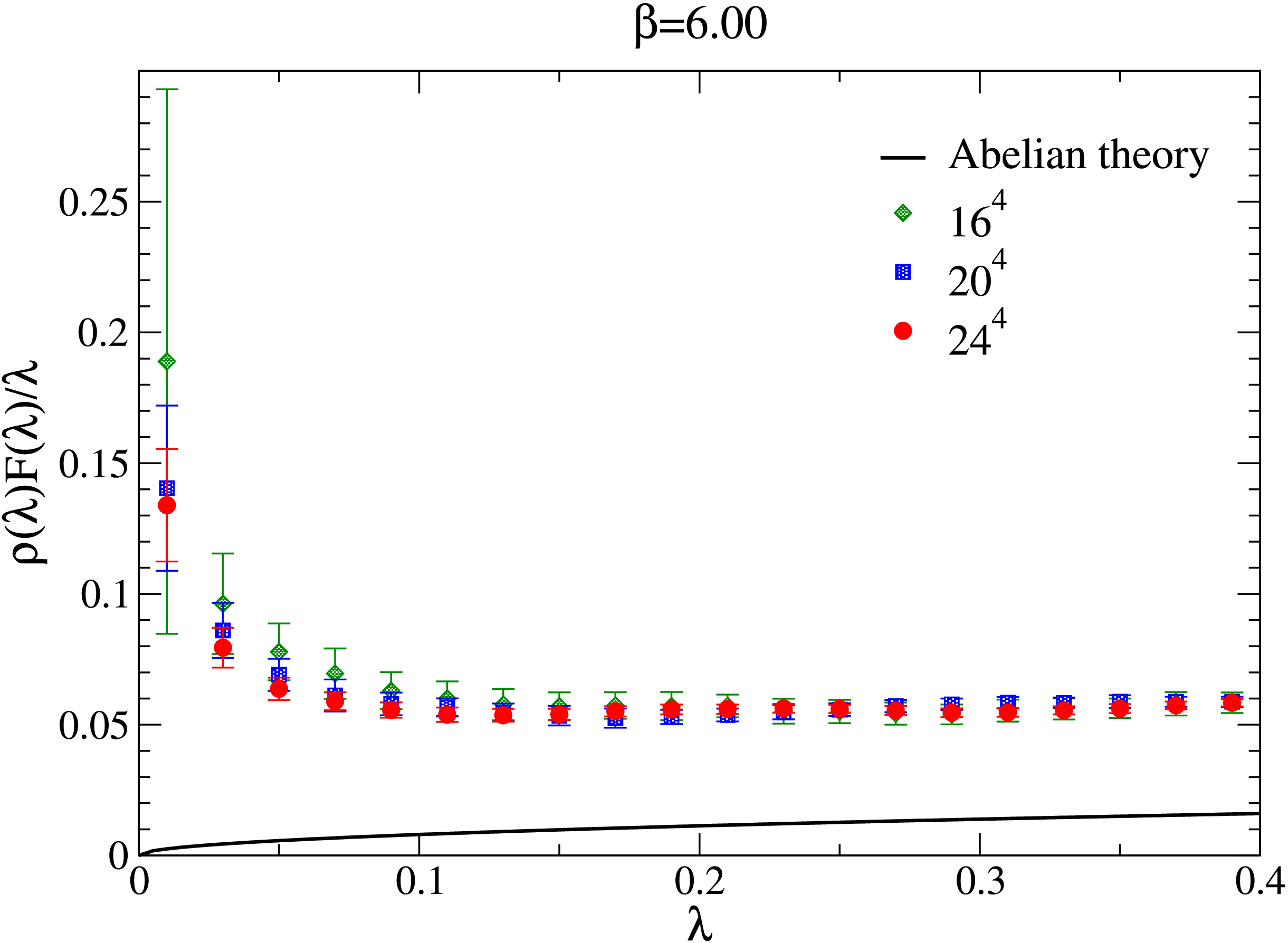}}
\caption{(a)  $\rho(\lambda)/\sqrt{\lambda}$ in the confinement phase. (b) $\rho(\lambda)F(\lambda)/\lambda$ vs. $\lambda$ in the confinement phase. The solid curve represents $\rho(\lambda)F(\lambda)/\lambda$ in the case of the abelian theory, $\rho(\lambda)F(\lambda)/\lambda=\sqrt{\lambda}/4\pi^2$.}
\label{FROW}
\end{center}
\end{figure}

To compare the behavior of $\rho(\lambda)$ in the non-abelian theory with that in the abelian theory, we plot $\rho(\lambda)/\sqrt{\lambda}$ in Fig. \ref{FROW}(a).
In the abelian theory it is constant because $\rho(\lambda)\sim\sqrt{\lambda}$ (see Eq. (\ref{abelian})).
In the non-abelian theory, we observe that $\rho(\lambda)/\sqrt{\lambda}$ is almost constant above $\lambda\sim 1$ [GeV].
In contrast, at small $\lambda$, $\rho(\lambda)/\sqrt{\lambda}$ is not constant and it seems to diverge as $\lambda\to 0$ in the infinite volume limit.
The FP eigenvalue density in the non-abelian theory shows a completely different behavior at small $\lambda$ compared to that of the abelian theory and we see the enhancement of the near-zero modes of the FP operator.
In the Gribov-Zwanziger scenario, these near-zero modes cause the color-Coulomb potential to be more singular in the infrared region than the simple pole. 

In Fig. \ref{FROW}(b) we plot $\rho(\lambda)F(\lambda)/\lambda$ as a function of $\lambda$.
As $\lambda$ approaches to 0, $\rho(\lambda)F(\lambda)/\lambda$ decreases for the free field (see Eq. (\ref{abelian})) while increases for the interacting field.
From this figure, we expect that $\rho(\lambda)F(\lambda)/\lambda$ diverges or goes to positive constant, and it is unlikely that $\rho(\lambda)F(\lambda)/\lambda$ goes to zero as $\lambda\to 0$ in the infinite volume limit.
Therefore, we conclude that the color-Coulomb self-energy of an isolated color charge is infrared divergent in SU(3) lattice gauge theory.

\subsection{$\langle\rho(\lambda)F(\lambda)\rangle/\lambda$ in the deconfinement phase}

\begin{figure}[htbp]
\begin{center}
\resizebox{7cm}{!}{\includegraphics{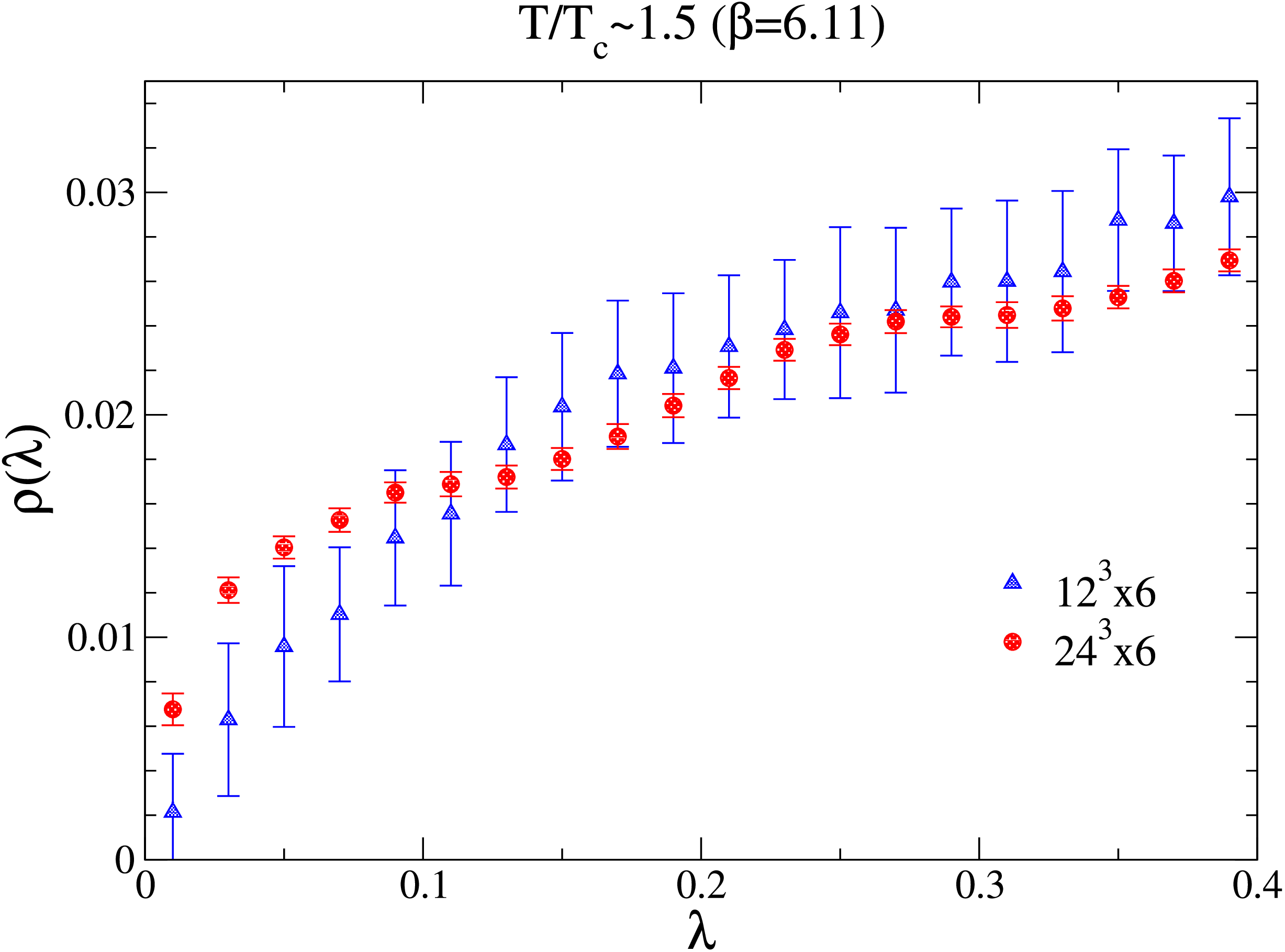}}
\resizebox{7cm}{!}{\includegraphics{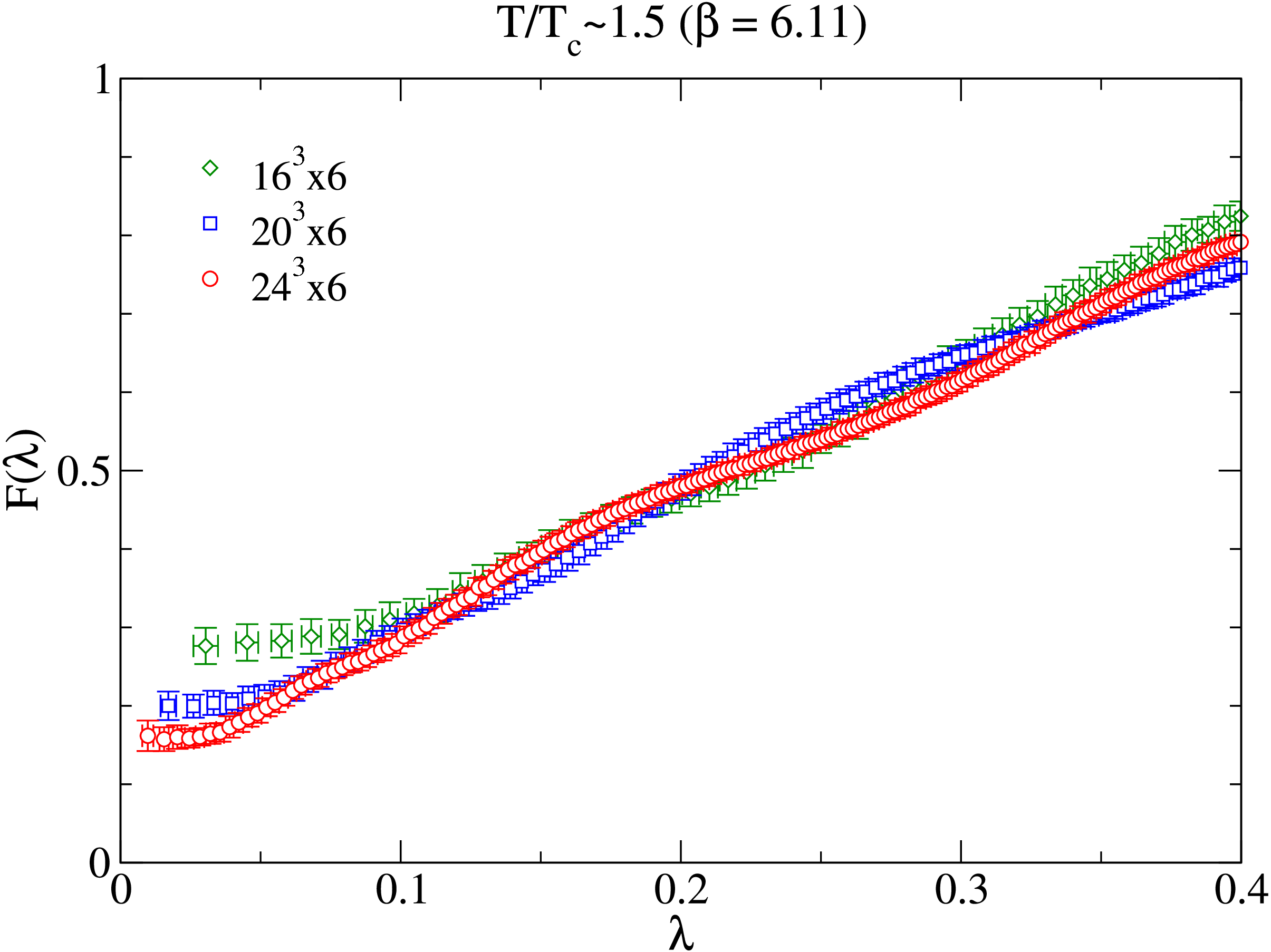}}
\resizebox{7cm}{!}{\includegraphics{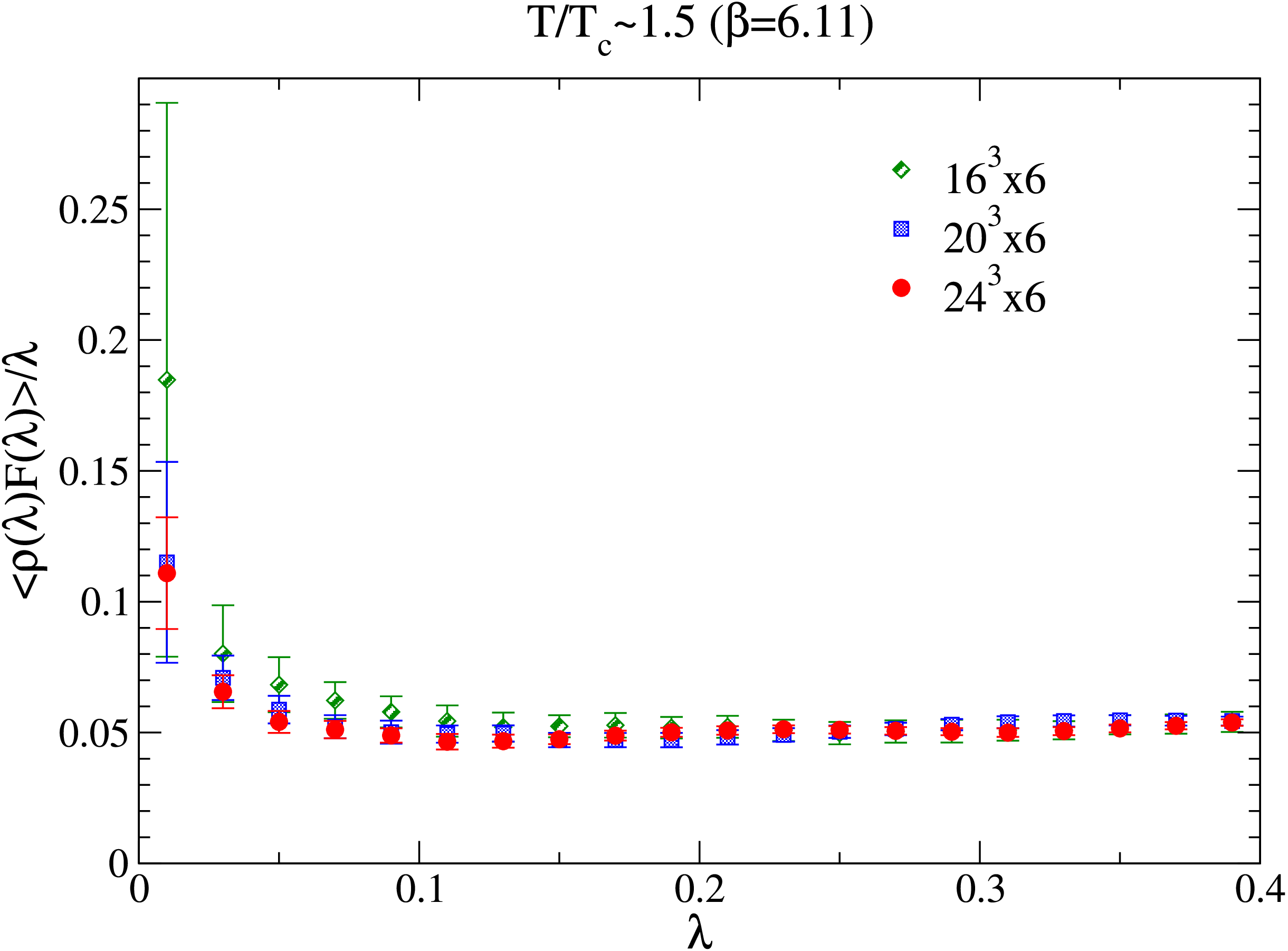}}
\caption{(a) The FP eigenvalue density in the deconfinement phase. (b) The average Laplacian $F(\lambda)$ in the deconfined phase. (c) $\rho(\lambda)F(\lambda)/\lambda$ vs. $\lambda$ in the deconfinement phase. }
\label{EVfinite}
\end{center}
\end{figure}

$\rho(\lambda)$, $F(\lambda)$ and $\rho(\lambda)F(\lambda)/\lambda$ at $T/T_c\sim 1.5$ ($T_c$ is the critical temperature of the phase transition) are displayed in Figs. \ref{EVfinite}(a), (b) and (c).
These figures show that there are no drastic changes of these behaviors in the deconfinement phase.
This is consistent with the fact that the color-Coulomb potential is not screened above the critical temperature.
We conclude that the necessary condition for color confinement is satisfied even in the deconfinement phase.




\section{Conclusions}

We have calculated the eigenvalue distribution of the FP operator in Coulomb gauge using quenched SU(3) lattice gauge simulations.
In the confinement phase, we observe the accumulation of the near-zero eigenvalues of the FP operator at large lattice volumes.
We conclude that the confinement criterion is satisfied in the SU(3) gauge theory.
Accordingly, the color-Coulomb potential becomes more singular than the simple pole in the infrared region.
This supports the Gribov-Zwanziger confinement scenario.
The results we obtained are qualitatively consistent with those of the SU(2) lattice simulation carried out by Greensite et al.

The near-zero modes of the FP operator survive above the critical temperature, and the behaviors of the FP eigenvalue density and the average Laplacian in the deconfinement phase are qualitatively the same as in the confinement phase.
Accordingly, the confinement criterion is satisfied even in the deconfinement phase in SU(3) gauge theory.
This would indicate that confining features survive even in the deconfinement phase, and we expect that further studies in Coulomb gauge provide insight into the understanding the strongly correlated quark-gluon plasma.

\section{Acknowledgments} 

The simulation was performed on an SX-5(NEC) vector-parallel computer at the RCNP of Osaka University. 
We appreciate the warm hospitality and support of the RCNP administrators.
This work is supported by Grants-in-Aid for Scientific Research from Monbu-Kagaku-sho (Nos. 13135216 and 17340080).

\bibliographystyle{h-physrev4}
\bibliography{forQCD}
\end{document}